\documentclass[conference]{IEEEtran}

\usepackage{cite}
\usepackage{amsmath,amssymb,amsfonts}
\usepackage{algorithmic}
\usepackage{graphicx}
\usepackage{textcomp}
\usepackage{xcolor}
\usepackage{url}
\usepackage{booktabs}
\usepackage[T1]{fontenc}

\graphicspath{{./images/}}

\newif\ifdoubleblindsubmission

\begin{document}
\bstctlcite{IEEEexample:BSTcontrol}

\title{Towards Online Malware Detection using Process Resource Utilization Metrics}

\author{
\IEEEauthorblockN{
Themistoklis Diamantopoulos\IEEEauthorrefmark{1},
Dimosthenis Natsos\IEEEauthorrefmark{1}\IEEEauthorrefmark{2},
Andreas L. Symeonidis\IEEEauthorrefmark{1}\IEEEauthorrefmark{2}
}

\IEEEauthorblockA{\IEEEauthorrefmark{1}Electrical and Computer Engineering Dept., Aristotle University of Thessaloniki, Thessaloniki, Greece}
\IEEEauthorblockA{\IEEEauthorrefmark{2}Cyclopt PC, Thessaloniki, Greece}
\IEEEauthorblockA{Emails: thdiaman@issel.ee.auth.gr, dcnatsos@ece.auth.gr, symeonid@ece.auth.gr}
}

\maketitle

\begin{abstract}
The rapid growth of Cloud Computing and Internet of Things (IoT) has significantly increased the interconnection of computational resources, creating an environment where malicious software (malware) can spread rapidly. To address this challenge, researchers are increasingly utilizing Machine Learning approaches to identify malware through behavioral (i.e. dynamic) cues. However, current approaches are limited by their reliance on large labeled datasets, fixed model training, and the assumption that a trained model remains effective over time-disregarding the ever-evolving sophistication of malware. As a result, they often fail to detect evolving malware attacks that adapt over time. This paper proposes an online learning approach for dynamic malware detection, that overcomes these limitations by incorporating temporal information to continuously update its models using behavioral features, specifically process resource utilization metrics. By doing so, the proposed models can incrementally adapt to emerging threats and detect zero-day malware effectively. Upon evaluating our approach against traditional batch algorithms, we find it effective in detecting zero-day malware. Moreover, we demonstrate its efficacy in scenarios with limited data availability, where traditional batch-based approaches often struggle to perform reliably.
\end{abstract}


\section{Introduction}
The era of Cloud Computing and Internet of Things (IoT) has accelerated the interconnection of computational resources, leading to a significant increase in malicious software (malware). Once it infiltrates a system, malware can compromise data, applications, or even the operating system \cite{Jansen:MalwareDefinition}. In response to this threat, several efforts have focused on identifying malware using machine learning methods.

Traditional detection approaches rely on either static or dynamic analysis. Static analysis examines a program's binary code to identify threats. Although generally effective, it often underperforms against code obfuscation and payload encryption techniques commonly employed to evade detection \cite{Aboaoja:MalwareChallengesOnline, NatsosSymeonidis:TransformersMalware}. Dynamic analysis, on the other hand, executes potentially malicious programs within controlled environments (e.g.\ sandboxes or virtual machines) to observe runtime behavior. It monitors system interactions, such as system calls, memory changes, and process resource utilization (CPU, memory, disk I/O, network activity) to uncover patterns invisible to code inspection.

Despite their effectiveness, dynamic analysis methods face key limitations. They rely on large labeled datasets, which are difficult to build and maintain \cite{NatsosSymeonidis:TransformersMalware, Bosinaki:TransferLearning}, and they often ignore the temporal evolution of malware. As the digital ecosystem evolves, driven by new software practices, shifting attack surfaces, and adaptive adversaries, malware continually changes its form through polymorphism, metamorphism, obfuscation, and structural variations designed to evade detection. This behavioral drift over time renders fixed models, trained on static or historical datasets, increasingly ineffective. Indeed, prior studies have shown that malware not only evolves rapidly in structure and behavior, but also increasingly mimics benign software to avoid detection, undermining the reliability of models trained on outdated data \cite{Jordaney:MalwareConceptDrift, Alenezi:MalwareEvolution}.

In this paper, we propose a methodology to address these limitations. We use a dataset in the domain of process resource utilization metrics and enrich it with temporal information indicating when each malware sample was introduced. We select this family of dynamic features because prior research has proven effective in detecting zero-day malware, even against sophisticated evasion techniques, due to their ability to capture cascading behavioral effects across tenant processes in operating systems \cite{NatsosSymeonidis:TransformersMalware, Abdelsalam:MalwareDataset}. Our approach enables assessment of how malware's temporal evolution affects detection performance over time, even with advanced dynamic detection techniques. To this end, we design an online learning methodology that adapts incrementally to emerging threats and illustrate how it can be more effective than traditional batch approaches in dynamic malware detection. Furthermore, we assess the effectiveness of our method in scenarios with scarce malware data, illustrating how it can still effectively detect malware in cases where batch approaches cannot perform reliably.

The rest of this paper is organized as follows. Section \ref{sec:relatedwork} reviews related work on dynamic malware detection. Section \ref{sec:methodology} describes our online malware detection approach. Section \ref{sec:evaluation} evaluates our approach, while Section \ref{sec:threatstovalidity} discusses our findings and indicates any threats to validity. Finally, Section \ref{sec:conclusion} concludes this work and provides insight for future work.

\section{Related Work} \label{sec:relatedwork}
Malware detection based on process resource utilization metrics has gained significant attention in recent years due to its ability to offer lightweight, platform-independent solutions for detecting malicious behavior. Some of the most commonly used metrics for this purpose include CPU usage, memory usage, disk I/O, and network activity. One of the major advantages of using these metrics is that they can be collected with minimal overhead, making them particularly suitable for systems with high scalability requirements, such as cloud environments or large enterprise networks.

The work presented in \cite{Abdelsalam:MalwareClustering} introduced a clustering-based approach for detecting malware infections in cloud virtual machines by analyzing resource utilization patterns. By modulating the sequential K-means algorithm to align with the hosted application architecture, the authors grouped VMs with similar behavior to identify infected systems through deviations from cluster centroids. Although promising, its effectiveness relies heavily on parameter tuning and expert knowledge, with limited success against low-profile malware that minimally impacts resources. Based on this, later research \cite{Abdelsalam:MalwareDataset} leveraged convolutional neural networks (CNNs) to classify VM processes as malicious or benign, based on resource utilization patterns. Incorporating time windows as additional dimensions improved detection accuracy and mitigated label inconsistencies, achieving an impressive 90\% detection rate.

A related study \cite{Tian:MalwareCNN} also employed CNNs to detect malware by combining runtime utilization data with memory object information. This dual-modal approach improved performance by integrating complementary features while using an out-grafting technique to retrieve process-level runtime data with minimal OS or hypervisor interaction. Conversely, \cite{Kimmel:MalwareRNN} adopted a recurrent neural network (RNN) based approach for online behavioral malware detection in cloud infrastructures. Unlike previous works that encoded input metrics as images, this study represented VM behavior as sequential performance metrics, including CPU, memory, and disk utilization. By comparing RNNs with CNNs, the authors demonstrated that RNNs are more robust to variations in input sequences, achieving superior overall performance. They also compared two RNN models (Bidirectional RNNs and Long Short-Term Memory RNNs), finding that LSTM RNNs provide a computationally efficient solution with comparable accuracy.

A different direction was taken in \cite{NatsosSymeonidis:TransformersMalware}, which applied transformer architectures to process resource utilization data. By encoding sequential process behavior (CPU, memory, disk usage), the authors improved detection accuracy while reducing dependence on large datasets. Their interpretability analysis further highlighted dynamic malware signatures and key behavioral features, even when the malicious process itself was not directly observable.

Though interesting, the aforementioned approaches are not incremental, meaning they are not updated over time and thus have limited effectiveness against zero-day malware. A few studies explore online malware detection but focus on different features. For example, studies \cite{Ceschin:OnlineMalwareStatic} and \cite{MishraStamp:OnlineMalwareStatic} investigate static online models, while \cite{FernandoKomninos:OnlineMalwareAPI} examines dynamic analysis at a lower level of abstraction, typically analyzing API calls as key indicators. Consequently, most research in this domain remains rooted in traditional feature sets. Online dynamic analysis methods based on process resource utilization metrics are still scarce. A notable contribution addressing this gap is \cite{Kegelmeyer:OnlineMalwareDetection}, which enables accurate and efficient real-time detection of malicious software. Its findings highlight the potential of stream-based methods to balance the trade-offs between batch and online learning in malware detection. Nevertheless, existing approaches continue to face several constraints: they depend on large labeled datasets that are costly to build and maintain, are typically trained once, making them prone to obsolescence as new threats emerge, and fail to account for malware's evolving nature, resulting in missed detections and false negatives. Their limited capacity to identify novel or zero-day variants exploiting unseen patterns exposes a persistent and fundamental weakness.

Our proposed approach addresses the limitations of existing methods by leveraging an online learning framework that adapts incrementally to emerging threats. By incorporating temporal information and extracting relevant features from process resource utilization metrics, our approach can adapt to long-term behavioral changes in malware; i.e.\ it can adapt to evolving malware behaviors over time due to shifting attack techniques, software environments, and adversarial tactics. This reflects our core hypothesis: that the performance of fixed detection models deteriorates over time as malware behavior changes in response to broader technological and strategic shifts. Furthermore, we demonstrate the efficacy of our method in scenarios with limited data availability, where traditional batch-based approaches often struggle to perform reliably.

\begin{figure*}[htbp]
\centerline{\includegraphics[width=\textwidth]{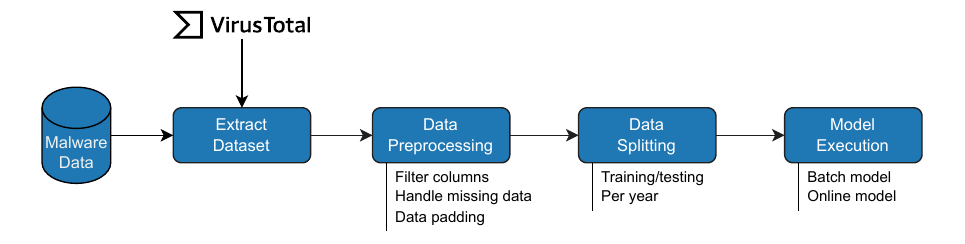}}
\caption{Steps of our online malware detection methodology\label{fig:architecture}}
\end{figure*}

\section{Online Malware Detection} \label{sec:methodology}
Our proposed approach is visualized in Figure \ref{fig:architecture}, which outlines the step-by-step process for creating a comprehensive malware detection framework. First, we utilize a publicly available dataset and supplement it with additional information from VirusTotal\footnote{\url{https://www.virustotal.com}} to create an enriched dataset. The data is then preprocessed through a series of steps, including filtering, handling missing values, and applying data padding techniques. Following this, the dataset is split in different ways to evaluate the performance of different scenarios. The preprocessed datasets are then fed into two models: a traditional batch-based and an online learning one. These components are further analyzed in subsequent sections, where we delve into the details of each step.

\subsection{Dataset Extraction}
\begin{figure}[htbp]
\centerline{\includegraphics[width=\columnwidth]{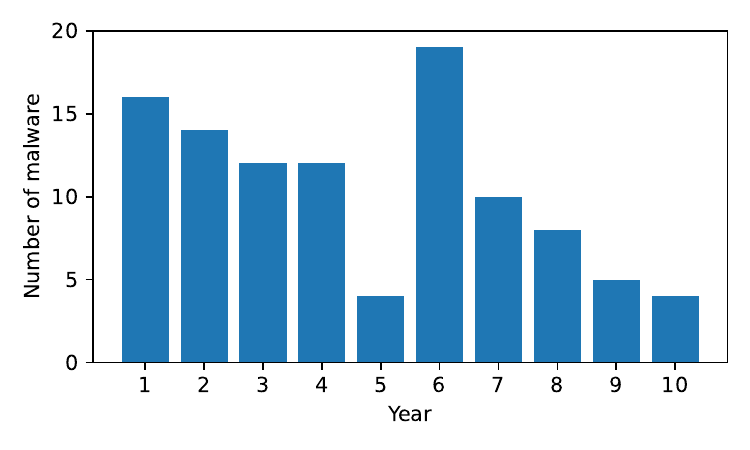}}
\caption{Number of malware per year}\label{fig:MalwarePerYear}
\end{figure}

To create a comprehensive dataset, we initially utilized the latest malware dataset \cite{Abdelsalam:MalwareDataset}, in the domain of process resource utilization metrics, which comprises features (such as CPU, Memory, Disk I/O usage, etc.) from various malware samples. Specifically, the dataset incorporates over 28,000 samples collected by monitoring the execution of 104 real malware variants (spanning DDoS/DoS, Backdoors, Trojans, Viruses, and Worms) each observed for one hour, with system-wide snapshots taken every 10 seconds. We enhance the dataset, by leveraging the VirusTotal API to gather the metadata of each malware sample, including its release year. By incorporating the year of introduction for each malware sample (which is commonly used feature \cite{MishraStamp:OnlineMalwareStatic}), we incorporate an essential temporal element that enables us to capture the evolution of malware over time. The resulted dataset comprised malware samples collected over a ten-year period. Figure \ref{fig:MalwarePerYear} illustrates the annual distribution of malware samples, covering the period from year one to year ten.

\subsection{Data Preprocessing}
To prepare the dataset for analysis, we first removed columns that act as identifiers (e.g.\ sample\_no, exp\_no, vm\_id, pid, ppid), represent execution metadata unavailable at inference time (e.g.\ sample\_time, process\_creation\_time), or are redundant with other behavioral metrics (e.g.\ num\_fds, label). Retaining such fields could introduce bias or information leakage without contributing to the behavioral characterization of malware. We then mapped the \textit{status} variable to a binary indicator, representing sleeping and running processes as 0 and 1, respectively. After ensuring that all remaining fields were numerical, any missing values were replaced with 0. Each dataset row corresponds to a process, and each column represents a metric. Table \ref{tab:features} summarizes all extracted features.

\begin{table}[!ht]
\caption{Features of the Malware Detection Dataset}
\label{tab:features}
\centering
\begin{tabular}{ll}
\toprule
Metric & Description\\
\midrule
cpu\_percent & Percentage of CPU usage\\
cpu\_num & Number of CPU cores allocated to process\\
cpu\_sys & CPU time spent in system-space\\
cpu\_user & CPU time spent in user-space\\
cpu\_children\_sys & CPU time (child processes) in system-space\\
cpu\_children\_user & CPU time (child processes) in user-space\\
num\_threads & Number of threads the process is using\\
mem\_shared & Memory shared with other processes\\
mem\_data & Memory of data (excluding code, shared)\\
mem\_vms & Virtual memory size used\\
mem\_rss & Memory used (code, data, shared)\\
mem\_dirty & Modified memory not written to disk\\
mem\_swap & Memory swapped to disk\\
mem\_lib & Memory used by shared libraries\\
mem\_uss & Unique memory used, excluding shared\\
mem\_text & Memory used for executable code\\
io\_write\_bytes & Total number of bytes written to disk\\
io\_read\_bytes & Total number of bytes read from disk\\
io\_write\_chars & Total number of characters written to disk\\
io\_read\_chars & Total number of characters read from disk\\
io\_write\_count & Number of write operations performed\\
io\_read\_count & Number of read operations performed\\
kb\_sent & Kilobytes of data sent over the network\\
kb\_received & Kilobytes of data received over the network\\
ionice\_ioclass & I/O scheduling class for the process\\
ionice\_value & I/O priority value for the process\\
nice & Process priority (adjusted for CPU time)\\
ctx\_switches\_voluntary & Number of voluntary context switches\\
ctx\_switches\_involuntary & Number of involuntary context switches\\
gid\_effective & Current process group ID\\
num\_fds & Number of open file descriptors\\
status & Process status (1 if running, 0 otherwise)\\
\bottomrule
\end{tabular}
\end{table}

We then flattened the dataset into a timestamp-based format, where each row captures the state of the entire system at a given time. Each row contains the concatenated feature vectors of all processes observed at that timestamp. To handle varying numbers of active processes per timestamp, we applied zero-padding so that all rows match the maximum observed dimensions: 227 processes and 32 metrics, resulting in 7264 columns. An additional column (inherited from the original dataset label) indicates whether the system is infected with malware at that timestamp. The final dataset consists of 28,213 samples, of which approximately 45\% (12,727) correspond to infected system states, resulting in a nearly balanced dataset.

Formally, for each timestamp $t$, our dataset comprises a sample of $m=227$ processes that may belong to one of two classes: the benign class (Equation \ref{eq:benign_sample}) and the infected class (Equation \ref{eq:infected_sample}). The former consists solely of benign processes ($bp$), while the latter also includes malicious processes ($mp$). Each process is represented by its feature vector ($f_1, f_2, \dots, f_n$) that comprises $n=32$ distinct features depicted in Table \ref{tab:features}. The aim of the experiments is to determine, given a sample at a specific timestamp, whether the system is infected with malware.

\begin{equation}
    \label{eq:benign_sample}
    X_t = 
    \underbrace{ f_1, f_2, \dots, f_n }_{\text{bp$_1$}},\ 
    \underbrace{ f_1, f_2, \dots, f_n }_{\text{bp$_2$}},\ 
    \dots,\ 
    \underbrace{ f_1, f_2, \dots, f_n }_{\text{bp$_m$}}
\end{equation}

\begin{equation}
    \label{eq:infected_sample}
    X_t = 
    \underbrace{ f_1, f_2, \dots, f_n }_{\text{bp$_1$}},\ 
    \underbrace{ f_1, f_2, \dots, f_n }_{\text{mp$_2$}},\ 
    \dots,\ 
    \underbrace{ f_1, f_2, \dots, f_n }_{\text{bp$_m$}}
\end{equation}

\subsection{Data Splitting}
To assess our methodology, we designed two data splitting scenarios. First, we compared the performance of our batch-based model against its online learning counterpart, where we split the data into training and test sets in a 60\%-40\% ratio randomly. This allowed us to assess how well each model performed under batch settings. After that, we designed experiments that mirror real-world scenarios, where our models are trained on data from year one and then tested on data from year two onwards, progressively updating the online model over time. By doing so, we simulated the online learning process, where our models continuously face new, evolving threats. By leveraging a decade-long dataset, we analyzed how the temporal evolution of malware impacts detection performance. As threats evolve, models trained on historical data gradually lose effectiveness, motivating adaptive mechanisms that can cope with emerging malware families over time. Importantly, the uneven distribution of malware samples across years (Figure \ref{fig:MalwarePerYear}) is preserved intentionally, as real-world malware prevalence is inherently low and irregular over time, and artificially balanced datasets may lead to misleading performance estimates \cite{Miranda:DebiasingAndroidMalware}.

\subsection{Model Execution} \label{subsec:modelexecution}
We implemented our batch model using Random Forests from scikit-learn \cite{Pedregosa:scikitlearn}, a popular machine learning library. We employed 10 base learners. Similarly, our online learning model utilized Adaptive Random Forests from CapyMOA \cite{Gomes:CapyMOA}, a library specifically designed for online learning. We also employed 10 base learners in this case. The key difference between the two implementations lies in the adaptive nature of the online model, which continuously updates its parameters based on new data that arrive, allowing it to adapt to evolving threat landscapes and improve its performance over time. Specifically, the model uses a so called test-then-train loop, meaning that instances are provided in chronological order one by one, the algorithm first predicts the target variable of each instance and then the instance is used for training.

\section{Evaluation} \label{sec:evaluation}
We perform three experiments, one to compare the performance of the online model and the batch model in a batch setting, one to compare them in an online setting, and one to assess the performance of the online model when the availability of ground-truth malware instances for training is limited.

\subsection{Batch Evaluation} \label{sec:batchevaluation}
\begin{figure}[htbp]
\centerline{\includegraphics[width=\columnwidth]{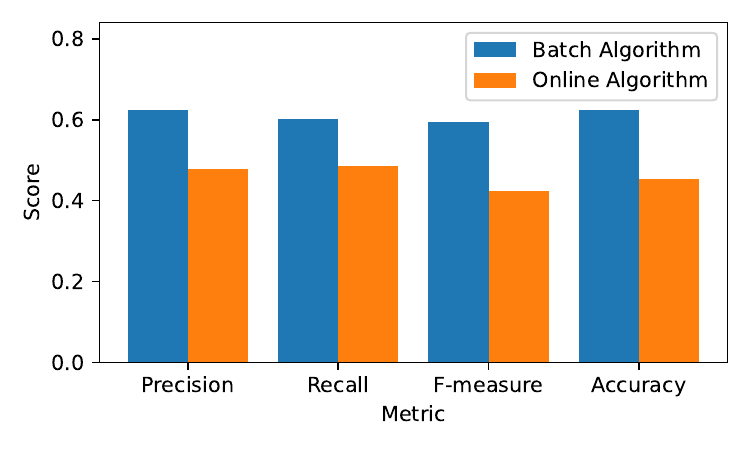}}
\caption{Evaluation of batch vs online model in a batch evaluation setting}\label{fig:BatchVsOnlineSplit}
\end{figure}

As already mentioned, in the batch evaluation experiment, we compared the performance of the Random Forest-based model with its online learning counterpart in a traditional batch learning scenario. That is, we split the dataset randomly into training and test sets (using a 60\%-40\% ratio). Both algorithms were trained on the training set (the online algorithm was trained one instance at a time) and run on the test set. The results are shown in Figure \ref{fig:BatchVsOnlineSplit}.

The results indicate that the batch algorithm outperforms the online learning approach in terms of all metrics when using a fixed training/testing split of 60/40. This is expected, as traditional batch-based models are optimized for fixed, labeled datasets, whereas online learning models need to adapt to new data in real-time. The online model's performance lags behind that of the batch algorithm due to its inability to leverage the full potential of the labeled training data. Interestingly, recall appears to be higher than precision for the online model, suggesting that while it is effective at detecting most malware processes, it may also produce more false positives.

\subsection{Online Evaluation} \label{sec:onlineevaluation}
\begin{figure}[htbp]
\centerline{\includegraphics[width=\columnwidth]{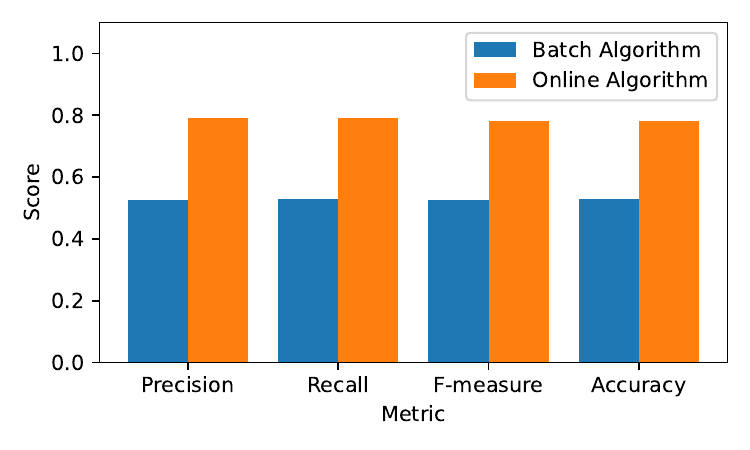}}
\caption{Evaluation of batch vs online model when the online model is updated}\label{fig:BatchVsOnlineAll}
\end{figure}

In contrast to the batch evaluation experiment, which involved a random dataset split, our second experiment assessed the performance of the online learning model when initially trained on data from year one and ran on data from year two onwards. This time, the online learning model was updated after predicting each instance of the test set (implementing the test-then-train loop described in subsection \ref{subsec:modelexecution}), while the batch model was trained once. Figure \ref{fig:BatchVsOnlineAll} depicts the evaluation metrics for the two models.
The results demonstrate that the online algorithm outperforms its batch-based counterpart, which is expected given the continuous adaptations made by the online model. Specifically, the online algorithm outperforms the batch-based one across all four evaluation metrics: Accuracy, Precision, Recall, and F-measure.

This is a particularly valuable result in the context of malware detection, where Recall, i.e.\ the proportion of actual malware instances correctly identified, is often prioritized over Precision or Accuracy. In security-sensitive domains, false negatives (i.e.\ undetected malware) can lead to severe consequences, whereas false positives are generally more tolerable. Therefore, models that achieve higher Recall, even at the cost of some additional false alarms, are typically preferred. This outcome highlights the advantage of using an online learning approach when working with datasets that are subject to frequent changes and updates, such as those in real-world malware scenarios. The online model's ability to learn from new data and adapt to evolving threat landscapes allows it to improve its performance over time, ultimately leading to higher effectiveness when compared to the batch algorithm.

\begin{figure}[htbp]
\centerline{\includegraphics[width=\columnwidth]{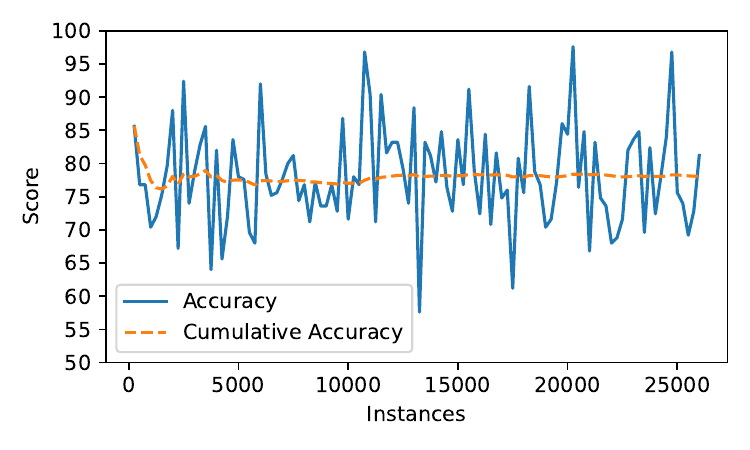}}
\caption{Accuracy of online model every 250 instances}\label{fig:OnlineAccuracy}
\end{figure}

Figure \ref{fig:OnlineAccuracy} further depicts the accuracy of the online model over time as it encounters new malware, examining its mean accuracy for every 250 instances and the cumulative accuracy (total over all instances already seen at any point in time).

The cumulative accuracy of the model is rather stable around 78\%, indicating that the online model is able to learn from each new sample and adjust its parameters accordingly, without experiencing a significant drop-off in performance. This also suggests that the online model is effectively capturing the underlying patterns and relationships in the data, even as it encounters new and previously unseen malware samples. Thus, this ability to maintain high accuracy over time highlights the benefits of using an online learning approach for real-world applications, where data is constantly being generated and updated. 

\subsection{Sample Malware Introduction} \label{sec:onlineevaluationrandomintro}
\begin{figure}[htbp]
\centerline{\includegraphics[width=\columnwidth]{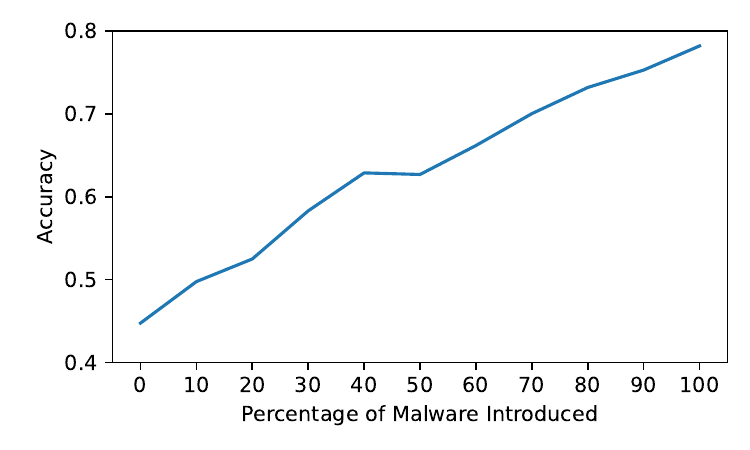}}
\caption{Evaluation of online model when a percentage of malware is introduced randomly}\label{fig:OnlineTrainingPercent}
\end{figure}

To further evaluate the robustness of our online learning model, we conducted a third experiment to account for the fact that not all malware samples may be known in order to be used for the model's ground truth. Thus, in this experiment, we maintain the same test data set (from year two onwards), however we only provide the malware target variable for training for a percentage of malware samples. We use different percentages from none (0\%) to all (100\%) malware. Practically, this limited data scenario simulates what would happen if our online model functioned in a realistic setting where it was updated only in cases where malware was detected (e.g.\ by security experts). In such a scenario, we cannot always assume that all malware would be found and given to the algorithm. The results are shown in Figure \ref{fig:OnlineTrainingPercent}.

Model performance improves as the proportion of malware instances used for training increases. For example, when roughly half of the dataset's malware processes are included, accuracy is about 65\%, whereas providing around 90\% raises accuracy to about 75\%. Furthermore, comparing Figure \ref{fig:BatchVsOnlineAll} and Figure \ref{fig:OnlineTrainingPercent} shows that even if less than a quarter of malware is identified by the security experts and supplied to the online model, it still outperforms the batch one. As expected, performance scales with the number of labeled malware samples, however the model can still achieve accurate results with only a subset of malware (e.g.\ about half or even less), emphasizing the value of online learning in practice.

\section{Discussion and Threats to Validity} \label{sec:threatstovalidity}
Our work demonstrates the effectiveness of online learning in dynamic malware detection, highlighting its potential as a robust and adaptive approach for real-world applications. However, several open issues and threats to validity warrant further exploration. One consideration is the dataset used in our experiments. While our study focuses on process resource utilization metrics, a well-established subset of dynamic malware analysis features, we acknowledge that broader validation across other dynamic data types (e.g.\ API calls, system calls, or memory dumps) would further demonstrate generalizability. The dataset of Abdelsalam et al.\ \cite{Abdelsalam:MalwareDataset} represents the current state of the practice for this feature family and was preprocessed following established methods in the literature \cite{Tian:MalwareCNN, NatsosSymeonidis:TransformersMalware}. Moreover, by training on earlier years and testing on later ones, our evaluation inherently captures the zero-day detection challenge, assessing the model's resilience to temporal drift.

The choice of models is also noteworthy, as recent advances in neural networks and transformers \cite{NatsosSymeonidis:TransformersMalware} have raised the bar for malware detection. Our approach focused on demonstrating the benefits of online learning rather than optimizing the underlying algorithm for accuracy. Future work should explore additional algorithms supported by the CapyMOA library and leverage its integration with PyTorch \cite{Paszke:PyTorch} to enhance model performance. Finally, we did not directly compare our results with existing batch learning approaches. Our goal was to assess the effectiveness of learning in an online setting, where such comparison is not directly applicable. To facilitate further research, we have made our code publicly available\footnote{\url{https://github.com/AuthEceSoftEng/online-malware-detection}} for anyone to reproduce our findings.

\section{Conclusion} \label{sec:conclusion}
In this paper, we proposed an online learning approach tailored to dynamic malware detection, specifically focusing on zero-day malware detection using process resource utilization metrics. By leveraging the capabilities of CapyMOA, our methodology enables the detection of zero-day malware in real-time, adapting to new and evolving threats as they emerge. 

The results of our evaluation demonstrate that our online learning approach outperforms traditional batch algorithms in detecting zero-day malware, with accuracy increasing significantly as more data is available. Furthermore, our methodology's ability to adapt to new threats and detect malware even when only a portion of the dataset is available indicates its potential value in realistic settings. 

As future work, we aim to further improve the effectiveness and robustness of our online learning approach, potentially exploring low-level dynamic features and different algorithms. Moreover, we plan to consider concept drifts, i.e.\ sudden changes in data distribution indicated by drops in accuracy, which can signify the emergence of new malware. 

\section*{Acknowledgment}
Parts of this work have been supported by the Horizon Europe project ECO-READY (Grant Agreement No 101084201), funded by the European Union.

\bibliographystyle{IEEEtran}
\bibliography{paper}

@IEEEtranBSTCTL{IEEEexample:BSTcontrol,
  CTLdash_repeated_names = "no"
}

@inproceedings{Abdelsalam:MalwareDataset,
  author = {Abdelsalam, Mahmoud and Krishnan, Ram and Sandhu, Ravi},
  title = {{Online Malware Detection in Cloud Auto-scaling Systems Using Shallow Convolutional Neural Networks}},
  year = {2019},
  isbn = {978-3-030-22478-3},
  publisher = {Springer-Verlag},
  address = {Berlin, Heidelberg},
  doi = {10.1007/978-3-030-22479-0_20},
  booktitle = {Proceedings of the 33rd Annual IFIP WG 11.3 Conference on Data and Applications Security and Privacy (DBSec 2019)},
  pages = {381--397},
  location = {Charleston, SC, USA}
}

@book{Jansen:MalwareDefinition,
  author = {Jansen, Wayne A. and Winograd, Theodore and Scarfone, Karen},
  title = {{Guidelines on Active Content and Mobile Code}},
  year = {2008},
  publisher = {National Institute of Standards and Technology},
  note = {NIST Special Publication 800-28 Revision 2},
  doi = {10.6028/NIST.SP.800-28ver2}
}

@article{NatsosSymeonidis:TransformersMalware,
  author = {Natsos, Dimosthenis and Symeonidis, Andreas L.},
  title = {{Transformer-based Malware Detection using Process Resource Utilization Metrics}},
  journal = {Results in Engineering},
  volume = {25},
  pages = {104250},
  year = {2025},
  issn = {2590-1230},
  doi = {10.1016/j.rineng.2025.104250}
}

@inproceedings{Bosinaki:TransferLearning,
  author = {Bosinaki, Konstantina and Natsos, Dimosthenis and Siachamis, Giorgos and Symeonidis, Andreas L.},
  title = {{From One Network to Another: Transfer Learning for IoT Malware Detection}},
  booktitle = {2025 IEEE International Conference on Cyber Security and Resilience (CSR)},
  year = {2025},
  pages = {287--294},
  doi = {10.1109/CSR64739.2025.11130005}
}

@article{Aboaoja:MalwareChallengesOnline,
  author = {Aboaoja, Faitouri A. and Zainal, Anazida and Ghaleb, Fuad A. and Al-rimy, Bander Ali Saleh and Eisa, Taiseer Abdalla Elfadil and Elnour, Asma Abbas Hassan},
  title = {{Malware Detection Issues, Challenges, and Future Directions: A Survey}},
  journal = {Applied Sciences},
  volume = {12},
  number = {17},
  pages = {8482},
  year = {2022},
  issn = {2076-3417},
  doi = {10.3390/app12178482}
}

@inproceedings{Abdelsalam:MalwareClustering,
  author = {Abdelsalam, Mahmoud and Krishnan, Ram and Sandhu, Ravi},
  title = {{Clustering-Based IaaS Cloud Monitoring}},
  booktitle = {2017 IEEE 10th International Conference on Cloud Computing (CLOUD)},
  year = {2017},
  pages = {672--679},
  doi = {10.1109/CLOUD.2017.90}
}

@article{Tian:MalwareCNN,
  author = {Tian, Donghai and Zhao, Runze and Ma, Rui and Jia, Xiaoqi and Shen, Qi and Hu, Changzhen and Liu, Wenmao},
  title = {{MDCD: A Malware Detection Approach in Cloud using Deep Learning}},
  journal = {Transactions on Emerging Telecommunications Technologies},
  volume = {33},
  number = {11},
  pages = {e4584},
  doi = {10.1002/ett.4584},
  year = {2022}
}

@article{Kimmel:MalwareRNN,
  author = {Kimmel, Jeffrey C. and Mcdole, Andrew D. and Abdelsalam, Mahmoud and Gupta, Maanak and Sandhu, Ravi},
  title = {{Recurrent Neural Networks Based Online Behavioural Malware Detection Techniques for Cloud Infrastructure}},
  journal = {IEEE Access},
  year = {2021},
  volume = {9},
  pages = {68066--68080},
  doi = {10.1109/ACCESS.2021.3077498}
}

@article{Ceschin:OnlineMalwareStatic,
  author = {Ceschin, Fabr\'{i}cio and Botacin, Marcus and Gomes, Heitor Murilo and Pinag\'{e}, Felipe and Oliveira, Luiz S. and Gr\'{e}gio, Andr\'{e}},
  title = {{Fast \& Furious: On the Modelling of Malware Detection as an Evolving Data Stream}},
  journal = {Expert Systems with Applications},
  volume = {212},
  pages = {118590},
  year = {2023},
  issn = {0957-4174},
  doi = {10.1016/j.eswa.2022.118590}
}

@inproceedings{Kegelmeyer:OnlineMalwareDetection,
  author = {Kegelmeyer, W. Philip and Chiang, Ken and Ingram, Joe},
  title = {{Streaming Malware Classification in the Presence of Concept Drift and Class Imbalance}},
  booktitle = {Proceedings of the 2013 12th International Conference on Machine Learning and Applications - Volume 02},
  year = {2013},
  pages = {48--53},
  isbn = {9780769551449},
  doi = {10.1109/ICMLA.2013.104},
  publisher = {IEEE},
  address = {USA},
}

@article{FernandoKomninos:OnlineMalwareAPI,
  author = {Fernando, Damien Warren and Komninos, Nikos},
  title = {{FeSA: Feature Selection Architecture for Ransomware Detection under Concept Drift}},
  journal = {Computers \& Security},
  volume = {116},
  pages = {102659},
  year = {2022},
  issn = {0167-4048},
  doi = {10.1016/j.cose.2022.102659}
}

@article{MishraStamp:OnlineMalwareStatic,
  author = {Mishra, Aniket and Stamp, Mark},
  title = {{Cluster Analysis and Concept Drift Detection in Malware}},
  journal = {Journal of Computer Virology and Hacking Techniques},
  year = {2025},
  volume = {21},
  number = {1},
  pages = {27--39},
  doi = {10.1007/s11416-025-00568-y},
  issn = {2263-8733}
}

@inproceedings{Gomes:CapyMOA,
  author = {Sun, Yibin and Gomes, Heitor Murilo and Lee, Anton and Gunasekara, Nuwan and Cassales, Guilherme Weigert and Liu, Justin Jia and Heyden, Marco and Cerqueira, Vitor and Bahri, Maroua and Koh, Yun Sing and Pfahringer, Bernhard and Bifet, Albert},
  title = {{Machine Learning for Data Streams with CapyMOA}},
  booktitle = {Machine Learning and Knowledge Discovery in Databases},
  year = {2026},
  pages = {438--443},
  publisher = {Springer Nature Switzerland},
  address = {Cham},
  isbn = {978-3-032-06129-4},
  doi = {10.1007/978-3-032-06129-4_27}
}

@article{Pedregosa:scikitlearn,
  author = {Pedregosa, F. and Varoquaux, G. and Gramfort, A. and Michel, V. and Thirion, B. and Grisel, O. and Blondel, M. and Prettenhofer, P. and Weiss, R. and Dubourg, V. and Vanderplas, J. and Passos, A. and Cournapeau, D. and Brucher, M. and Perrot, M. and Duchesnay, E.},
  title = {{Scikit-learn: Machine Learning in Python}},
  journal = {Journal of Machine Learning Research},
  volume = {12},
  pages = {2825--2830},
  year = {2011}
}

@inproceedings{Paszke:PyTorch,
  author = {Paszke, Adam and Gross, Sam and Massa, Francisco and Lerer, Adam and Bradbury, James and Chanan, Gregory and Killeen, Trevor and Lin, Zeming and Gimelshein, Natalia and Antiga, Luca and Desmaison, Alban and K\"{o}pf, Andreas and Yang, Edward and DeVito, Zach and Raison, Martin and Tejani, Alykhan and Chilamkurthy, Sasank and Steiner, Benoit and Fang, Lu and Bai, Junjie and Chintala, Soumith},
  title = {{PyTorch: an imperative style, high-performance deep learning library}},
  booktitle = {Proceedings of the 33rd International Conference on Neural Information Processing Systems},
  year = {2019},
  pages = {721},
  doi = {10.5555/3454287.3455008},
  publisher = {Curran Associates Inc.},
  address = {Red Hook, NY, USA}
}

@inproceedings{Jordaney:MalwareConceptDrift,
  author = {Jordaney, Roberto and Sharad, Kumar and Dash, Santanu Kumar and Wang, Zhi and Papini, Davide and Nouretdinov, Ilia and Cavallaro, Lorenzo},
  title = {{Transcend: detecting concept drift in malware classification models}},
  booktitle = {Proceedings of the 26th USENIX Conference on Security Symposium},
  year = {2017},
  pages = {625--642},
  location = {Vancouver, BC, Canada},
  publisher = {USENIX Association},
  address = {USA}
}

@article{Alenezi:MalwareEvolution,
  author = {Alenezi, Mohammed N. and Alabdulrazzaq, Haneen and Alshaher, Abdullah A. and Alkharang, Mubarak M.},
  title = {{Evolution of Malware Threats and Techniques: A Review}},
  journal = {International Journal of Communication Networks and Information Security},
  volume = {12},
  number = {3},
  pages = {326--337},
  year = {2020},
  issn = {2073-607X},
  publisher = {Kohat University of Science and Technology (KUST)}
}

@article{Miranda:DebiasingAndroidMalware,
  author = {Miranda, Tom\'as Concepci\'on and Gimenez, Pierre-Fran\c{c}ois and Lalande, Jean-Fran\c{c}ois and Tong, Val\'erie Viet Triem and Wilke, Pierre},
  title = {{Debiasing Android Malware Datasets: How Can I Trust Your Results If Your Dataset Is Biased?}},
  journal = {IEEE Transactions on Information Forensics and Security},
  volume = {17},
  pages = {2182--2197},
  year = {2022},
  doi = {10.1109/TIFS.2022.3180184}
}

\end{document}